\documentclass[preprint,sort&compress,12pt]{elsarticle}

\usepackage[top=1in, bottom=1in, left=1in, right=1in]{geometry}

\usepackage{graphicx}
\usepackage{amsmath}
\usepackage{epstopdf}
\usepackage{natbib}
\usepackage{amssymb}

\usepackage{amsthm}

\usepackage{lineno}
\usepackage{subfig}
\usepackage{enumerate}
\usepackage{algorithm}
\usepackage{algpseudocode}
\usepackage{hyperref}
\usepackage{fixltx2e}

\usepackage{lipsum}
\usepackage{fancyhdr}
\journal{ASCE-ASME}

\begin{document}

\begin{frontmatter}




  \title{Propagation of Input Uncertainty in Presence of Model-Form Uncertainty: A Multi-fidelity
    Approach for CFD Applications}



\author{Jian-Xun Wang}
\author{Christopher J. Roy}
\author{Heng Xiao\corref{corxh}}
\ead{hengxiao@vt.edu}
\cortext[corxh]{Corresponding author. Tel: +1 540 231 0926}

\address{Department of Aerospace and Ocean Engineering, Virginia Tech, Blacksburg, VA 24060, United States}

\begin{abstract}
Proper quantification and propagation of uncertainties in computational simulations are of critical importance. This issue is especially challenging for computational fluid dynamics (CFD) applications.  A particular obstacle for uncertainty quantifications in CFD problems is the large model discrepancies associated with the CFD models used for uncertainty propagation. Neglecting or improperly representing the model discrepancies leads to inaccurate and distorted uncertainty distribution for the Quantities of Interest.  High-fidelity models, being accurate yet expensive, can accommodate only a small ensemble of simulations and thus lead to large interpolation errors and/or sampling errors; low-fidelity models can propagate a large ensemble, but can introduce large modeling errors. In this work, we propose a multi-model strategy to account for the influences of model discrepancies in uncertainty propagation and to reduce their impact on the predictions. Specifically, we take advantage of CFD models of multiple fidelities to estimate the model discrepancies associated with the lower-fidelity model in the parameter space. A Gaussian process is adopted to construct the model discrepancy function, and a Bayesian approach is used to infer the discrepancies and corresponding uncertainties in the regions of the parameter space where the high-fidelity simulations are not performed. Several examples of relevance to CFD applications are performed to demonstrate the merits of the proposed strategy. Simulation results suggest that, by combining low- and high-fidelity models, the proposed approach produces better results than what either model can achieve individually.
\end{abstract}

\begin{keyword}
Uncertainty\sep Probability\sep Aerospace
\end{keyword}

\end{frontmatter}


\section{Introduction}
\label{S:Intro}
Computational simulations, particularly those based on numerical solutions of 
partial differential equations, have become an indispensable tool for decision-makers 
by providing predictions of system responses with quantified uncertainties.  It is 
important yet challenging to quantify these uncertainties to provide reliable guidance for 
decision-making processes. When a physical system is simulated with a model, 
uncertainties may result from various sources, which can be broadly classified into 
data uncertainties and model uncertainties~\cite{LeMaitre:2010vg}. Data uncertainties 
are due to intrinsic variations of the system itself or inexact characterizations thereof, 
including operation scenarios, initial and boundary conditions, geometry, and material 
properties, among others. As such, they are also known as \emph{input uncertainties} 
or \emph{parametric uncertainties}. Throughout this paper we use the flow past an airfoil 
as an example, which is a classical problem in Computational Fluid Dynamics (CFD). In 
this example, input uncertainties may originate from (1) operation conditions such as angle of 
attack (AoA), Reynolds number, and Mach number, (2) initial and boundary conditions 
of the flow field, (3) geometry of the airfoils, and (4) material properties (e.g., roughness 
of the airfoil surface), among others. We assume that the input uncertainties have 
already been adequately characterized and quantitatively represented with known 
probability distributions. The quantified input uncertainties are then propagated into 
predictions of the quantities of interest (QoIs) through the CFD models, leading to 
uncertainties in the output. However, the output uncertainty distribution can be distorted
by the uncertainties stemmed from computational models. Model uncertainty or model 
discrepancy, which refers to the differences between the mathematical model and the 
physical reality, is a main factor affecting uncertainty propagation. Interpolation error is 
another factor that may affect input uncertainty propagations when the simulations are 
too expensive to allow for a statistically meaningful number of samples to be propagated 
and thus surrogate models are used. The discrepancy between the original expensive 
model output and the relatively cheap surrogate model output is referred to as interpolation 
error. There are other uncertainties associated with the numerical model, e.g., from 
parameter calibration and numerical discretization, which are not of particular concern here.
In CFD applications, perfect models are rarely available or affordable for any nontrivial 
turbulent flows, and the discrepancies between model predictions and physical system 
responses are often appreciable. These model discrepancies distort the output uncertainties 
and complicate the uncertainty propagation processes. High-fidelity models of turbulent 
flows, being accurate (if used properly) yet expensive, can accommodate only a small 
ensemble of simulations and thus lead to large interpolation errors and/or sampling 
errors; low-fidelity models can propagate a large ensemble, but can introduce large
modeling errors (e.g., when Reynolds-Averaged Navier--Stokes (RANS) model is used 
on flows with massive separation or strong pressure gradient). Therefore, multiple models 
with different fidelities are employed in this work in a complementary way to address this 
difficulty.

The benefits of utilizing computational models of multiple fidelities have long been recognized 
by the science, engineering, and statistics communities. Various statistical frameworks have 
been proposed to incorporate information from multiple models. Bayesian Model Averaging 
(BMA) method~\cite{Hoeting:1999vk} and its dynamic variant~\cite{Raftery:2010je} compute 
the distributions of the prediction by averaging the posterior distribution under each of the 
models considered, weighted by their posterior model probability. However, the fundamental 
assumption in BMA is that the models of concern are all plausible, competitive explanations 
of the data, and there is no hierarchy of these models. While this context may be relevant 
for certain applications, in engineering we often have a well-defined hierarchy of models from 
cheap, low-fidelity models to expensive, high-fidelity models, and thus BMA methods are
not of particular interest in the context of uncertainty quantification in CFD problems. In 
contrast to BMA methods, for which a model hierarchy does not exist, Multi-Level Monte Carlo
(MLMC) methods create and utilize such a hierarchy to facilitate uncertainty propagation by running
the same simulator on a series of successively refined meshes, leading to a sequence of multi-level
models.  By optimizing the number of simulations conducted on each level, significant speedup
compared to the standard MC method has been obtained~\cite{Mishra:2012ec}. 
However, the MLMC method in essence is a control variate based \emph{variance reduction} 
technique with two critical assumptions~\cite{Giles:2008gc}: (1) the ensembles computed on 
different levels are all \emph{unbiased} estimators of the QoI, and (2) the variance of the 
estimator decreases as the mesh is refined. These assumptions are also used in the multi-model
extension~\cite{muller2014solver}. Therefore, model discrepancies are not accounted for in MLMC
methods. In turbulent flows applications, however, model discrepancies play a pivotal role, and thus
the MLMC method is less likely to be applicable, since the assumptions above cannot be fully
justified. Data assimilation represents another approach of combining multiple sources of 
information. Recently, Narayan et al.~\cite{Narayan12} proposed a framework to assimilate
multiple models and observations, which expresses the assimilated state as a general linear
functional of the models and the data. In general, model discrepancies are not explicitly considered in 
data assimilation methods, although the available data may implicitly correct any biases in the model
predictions in the filtering operations.  Similar to MLMC methods, the filtering operations in data
assimilations aim to reduce variances of the overall estimation from several unbiased sources. 

Gaussian process (GP) is a promising approach to account for model discrepancies. The basic
assumption is that the quantities of interest at different design points obey multivariate joint
Gaussian distributions. With this assumed correlation, Bayesian updating can be used to infer the
distribution of the QoI in the entire field from the data available at certain locations. Results 
obtained from the Bayesian inference can be viewed as a probabilistic response surface, i.e., mean predictions with quantified uncertainties. Gaussian processes are routinely used to construct the prior in geostatistic
and atmospheric inverse problems~\cite{cressie1993statistics,gourdji2008global,rodgers2000inverse}. 
In engineering design optimization it is used to build surrogate models based on simulations performed 
at a small number of points in the parameter space~\cite{forrester2007multi,xiong2008new}. 
Kennedy and O'Hagan~\cite{kennedy2000pre} developed the first
multi-model prediction framework accounting for model discrepancies by using Gaussian process 
and Bayesian inference. An important assumption of their framework is that the predictions from 
different levels are auto-regressive. That is, the higher-level model prediction $z^H$ is related to 
the lower level model prediction $z^L$ by a regressive coefficient $\rho$ and a discrepancy $\delta$, 
i.e., $z^H =\rho z^L + \delta$, where $\delta$ is described by a GP.  Another assumption is that the 
prior for the prediction of each model can be described by stationary GP.  Simulation data obtained 
from all models are then used to infer the predictions of the highest-fidelity model. More than a decade
after being proposed, although this multi-fidelity modeling framework has been used and developed
in mechanical and optimizations~\cite{xiong2008new,park2016remarks}, it has yet to find 
widespread use for the CFD applications, and thus much more investigations are still needed. 
 
The objective of this work is to account for model discrepancy in the propagation of input
uncertainties in CFD applications.  Information from models of multiple fidelities is used to
enable efficient propagation of input uncertainties. In fluid dynamic problems, the low-fidelity 
models (e.g., panel methods) are relatively cheap computationally, with each evaluation taking 
only seconds or minutes. Consequently, the input parameter space can be sampled sufficiently 
when performing low-fidelity simulations, and thus Gaussian process modeling of the interpolation 
errors for the low-fidelity predictions as conducted in Kennedy and O'Hagan~\cite{kennedy2000pre}
is largely unnecessary. Therefore, we propose a simpler method based on the framework of~\cite{kennedy2000pre} and classical Bayesian approach. Specifically, we describe only the model discrepancy $\delta$ between the low- and high-fidelity models ($z^L$ and $z^H$, respectively) with
Gaussian processes without assuming the responses themselves to have a certain distribution. 
We propagate a large ensemble $\mathcal{D}^L$ by using a low-fidelity model $z^{L}$, and then 
use realizations of model discrepancies (i.e., random draws from the fitted Gaussian 
process $\mathcal{GP}$) to correct $z^L$. With this strategy the distortion caused by low-fidelity model
deficiency on the output uncertainty distribution can be reduced with the information provided by 
the high-fidelity simulations. Previous authors have used GPs and multi-model framework to improve
predictions~\cite[e.g.,][]{higdon2004co,le2013bayesian} and to serve as surrogate models in design
optimizations~\cite[e.g.,][]{forrester2007multi,xiong2008new,huang2006sequential,zhou2016active}. 
However, to the authors' knowledge, the use of GPs and multi-fidelity models to account for model 
discrepancies and to perform implicit model calibrations in uncertainty propagation for CFD 
applications is a novel development. The coupling between model uncertainties and input uncertainties is a unique issue in using GP-based probabilistic response surface for uncertainty propagation, and this challenge is not present when using GP for predictions or optimizations.
This issue will be discussed in detail with the complete synthetic cases in the electronic supplementary materials.
 
The rest of the paper is organized as follows. The methodology and algorithm of the multi-model 
uncertainty propagation scheme are presented in Section~\ref{S:Meth}. Test cases are investigated 
to demonstrate the proposed method, and the results are presented in Section~\ref{S:Test}. The 
comparison of the multi-model strategy with traditional single-model approaches are discussed in Section~\ref{S:diss}. Finally, Section~\ref{S:Con} concludes the paper.

\section{Methodology}
\label{S:Meth}
An essential part of the proposed uncertainty propagation method is developed based on 
the multi-model framework of~\cite{kennedy2000pre}. Specifically, we use the high-fidelity 
simulations performed on a small ensemble, along with the low-fidelity simulations on the 
same ensemble, to construct a probabilistic response surface for the model discrepancy~$\delta$. 
Samples drawn from the probabilistic response surface are then used to correct the low-fidelity 
simulations, which can be performed on an ensemble as large as required. The correction 
of the low-fidelity results by using data from the high-fidelity simulations can be considered 
implicit model calibration. In this section, we first introduce Gaussian process modeling and 
Bayesian inference of model discrepancy in Section~\ref{S:gp}. The overall algorithm of the 
multi-model uncertainty propagation is then presented in Section~\ref{S:algo}.

\subsection{Gaussian Process Modeling and Bayesian Inference
 of Model Discrepancy}
\label{S:gp}
Referring to the airfoil example again, the model discrepancy $\delta$ is approximated by the
difference between the lift coefficients predicted by the low- and high-fidelity models, and its
magnitude depends on the AoA, denoted as input variable $x$.  Since it is prohibitively expensive to
perform high-fidelity simulations of the flow over an airfoil for a large number of AoA, the
discrepancy $\delta$ is effectively an unknown function of $x$. To facilitate modeling, we assume that
(1) the discrepancy $\delta(x_i)$ at any location $x_i$ is a random variable obeying a Gaussian
distribution, and (2) the discrepancies $\delta(x_i)$, where $i = 1, \cdots, n$, at any number
of $n$ locations have a joint Gaussian distribution.  Consequently, the unknown function $\delta(x)$
is a Gaussian process, which is expressed as~\cite{williams2006gaussian}:
\begin{equation} 
\label{E:GP}
\delta(x) \sim \mathcal{GP}(m, k),
\end{equation} 
where the mean function $m(x)$ is the expectation of $\delta(x)$; the covariance or kernel function
$k(x, x')$ dictates the covariance between the values of function $\delta$ at two locations $x$ and
$x'$.

In this work it is assumed that the low-fidelity model is explicitly calibrated beforehand with the best
available information (but not against high-fidelity simulation results) to eliminate any bias over
the entire input parameter domain, and thus a zero mean function $m(x) =0$ is used. Note that the
zero mean function represents the prior of model discrepancy without information from high-fidelity results and can be corrected by incorporating high-fidelity data through Bayesian updating. 
We also can assign the prior $m(x)$ with a polynomial function, but it means that more hyperparameters 
are to be inferred, which requires more data from the expensive high-fidelity CFD simulations.
Assuming $\delta(x)$ is a smooth function over most part of the parameter domain, the following 
kernel function is chosen, i.e.,
\begin{equation}
\label{E:kF}
k (x, x') =  \sigma^2_f \exp 
\left( 
  \dfrac{-|x - x'|^2}{2l^2} 
\right)
\end{equation}
where $| \cdot |$ denotes Euclidean norm, $\sigma_f$ determines the overall magnitude of the
variance, and $l$ is the length scale. $\sigma_f$ and $l$ are referred to as hyperparameters. This
kernel function is stationary in that the covariance between two points only depends on their
distances and not on their specific locations. While this may not be a good assumption in some cases, 
it can be a reasonable choice when only a small amount of data are available, and it is thus adopted in 
the present study. More sophisticated, possibly non-stationary, kernel functions are 
possible~\cite{williams2006gaussian,gramacy2008bayesian}, but this
issue is beyond the scope of this work.

If at certain locations $\mathbf{x}_o = [ x^{o}_1, \cdots, x^{o}_{nobs} ]^T$ in the parameter space,
the values of the discrepancy are $\mathbf{y}_o = [\delta^o_1, \cdots, \delta^o_{nobs}]$, the
objective is to predict the values $\mathbf{y}_p$ of the function $\delta$ at some other locations
$\mathbf{x}_p = [x^p_1, \cdots, x^p_{npred}]^T$.  Subscripts $o$ and $p$ indicate observations and
predictions, respectively. Observations in this work refer to the process of performing low- and
high-fidelity simulations to ``observe'' the values of the discrepancy function at locations
$\mathbf{x}_o$.

Following Arendt et al.~\cite{arendt2012quantification}, a modular approach is adopted here to 
determine the hyperparameters and the posterior predictions. Specifically, the hyperparameters 
are first inferred from observation $\mathbf{y}_o$ via Maximum Likelihood Estimation (MLE) before the predictions $\mathbf{y}_p$ are
inferred. The hyperparameter pair $\sigma_f$ and $l$ that maximizes the likelihood of
obtaining observation $y_o$ is chosen. In practice, the logarithm of the likelihood $p$ of
$\mathbf{y}_o$ conditioned on hyperparameters, i.e.,
\begin{equation}
  \label{eq:loglike}
  \log p (\mathbf{y}_o|\mathbf{x}_o, \sigma_f, l) 
  = 
  -\frac{1}{2} \mathbf{y}_o^T K_o \mathbf{y}_o 
  -\frac{1}{2} \log \det(K_o)
  -\frac{n_{obs}}{2}  \log (2 \pi)
\end{equation}
is maximized, $\det(K_o)$ is the determinant of matrix $K_o$. The optimization is performed with 
standard gradient-based procedure~\cite{williams2006gaussian}. Once chosen, the hyperparameters 
are considered fixed in subsequent inference of $\mathbf{y}_p$.

Before the values of $\mathbf{y}_o$ are known, the elements in the stacked vector $[\mathbf{y}_o,
\mathbf{y}_p]^T$ have a joint Gaussian distribution according to the definition of GP,
\begin{equation} 
\label{E:priorGP}
	\left[\!
	\begin{array}{c}
	\mathbf{y}_o\\
	\mathbf{y}_p
	\end{array}
	\!\right] \sim 
	\mathcal{N}
	\left( 
	\left[\!
	\begin{array}{c}
	\mathbf{0}\\
	\mathbf{0}
	\end{array}
	\!\right],
	\left[\!
	\begin{array}{cc}
	K_o & K_{op}\\
	K_{op}^T & K_p
	\end{array}
	\!\right]
	\right) ,
\end{equation}
where $\mathcal{N}$ denotes Gaussian distribution.  The mean vector $[\mathbf{0}, \mathbf{0}]^T$ is
the value of mean function $m(x)$, chosen to be zero, evaluated at $[\mathbf{x}_o, \mathbf{x}_p]^T$.
The covariance matrix $K$ is obtained in a similar way by evaluating the kernel function $k$, and it
is partitioned to sub-matrices $K_o$, $K_p$ and $K_{op}$, corresponding to the auto-variances of
$y_o$ and $y_p$, and their cross-covariance, respectively. When $\mathbf{y}_o$ is known, the
distribution of $\mathbf{y}_p$ conditional on $\mathbf{y}_o$ is still a joint Gaussian, which can be
obtained with standard Bayesian inference procedure as the following~\cite{williams2006gaussian}:
\begin{equation}
\label{E:postGP}
\mathbf{y}_p|\mathbf{y}_o, \sigma_f, l \sim 
\mathcal{N} (K_{op}^TK_o^{-1}\mathbf{y}_o,  \; K_p -  K_{op}^TK_o^{-1}K_{op})  ,
\end{equation}
which is the posterior distribution of $\mathbf{y}_p$. In contrast, Eq.~(\ref{E:priorGP}) specifies
the prior distribution of $\mathbf{y}_p$. The changes in the mean and the covariance of the posterior distribution reflect the new information provided by the observations $\mathbf{y}_o$.

\subsection{Multi-Model Uncertainty Quantification Method}
\label{S:algo}
After detailing the procedure of constructing a probabilistic response surface for the model
discrepancy above, which is at the core of current multi-model method, we outline below the 
overall algorithm of the uncertainty propagation referring to the airfoil example above.

Given the joint uncertainty distribution of input $x$, e.g., the operation conditions (AoA and
Mach number), the objective is to find the uncertainty distribution of the QoI, e.g., the lift
coefficient.  Assuming, again, that the input uncertainties have been characterized with known
probabilities, a traditional MC uncertainty propagation procedure consists of the following three
steps:
\begin{description}
\item [Sampling:] Sample the input uncertainty with a space filling method (e.g., the Latin
  Hypercube sampling~\cite{Helton:2003fc}) to obtain an ensemble $\mathcal{D}^L = \{x_j\}_{j =
    1}^{N^L}$, where $j = 1, \cdots, N^L$, and $N^L$ is the number of samples in $\mathcal{D}^L$.
\item [Propagation:] Propagate each sample $x_j$ in $\mathcal{D}^L$ with a model $z^L$, i.e., ${f}_j
  = z^L(x_j)$, leading to an output ensemble $\mathcal{F}^L = \{f_j\}_{j = 1}^{N^L}$.
\item [Aggregation:] Aggregate the output ensemble $\mathcal{F}^L$ to obtain the output uncertainty
  distribution of the QoI.
\end{description}
If the responses given by the numerical model $z^L$ differ from the true response, the obtained QoI
uncertainty distribution would be distorted. Increasing the fidelity of the model used in the propagation 
step above would generally reduce the model discrepancy and thus decrease the distortion in the output 
uncertainty. However, high-fidelity models have high computational costs and would reduce the number 
of samples that can be afforded, leading to increased sampling errors. With given computational 
resources, increasing the number of samples and increasing the model fidelity contradict each other by 
competing for resources.

In view of these difficulties, we perform high-fidelity simulations on a small number $N^H$ of samples 
in $\mathcal{D}^H$ to obtain an ensemble $\mathcal{F}^H$. Assuming a GP for the prior of the 
model discrepancy $\delta$ as in Eq.~(\ref{E:priorGP}), the simulation results in ensemble $\mathcal{F}^H$ 
are then used to obtain the posterior distribution according to Eq.~(\ref{E:postGP}).  A number $N_c$ of 
realizations of the discrepancy function are drawn from the posterior distribution to correct the 
output ensemble $\mathcal{F}^L$ given by the low-fidelity model, with each realization corresponding 
a corrected ensemble. The corrections lead to a set of $N_c$ ensembles, which is denoted 
as $\{\hat{\mathcal{F}}_k^L\}_{k = 1}^{N_c}$.  In this case each element in the set is an 
ensemble $\mathcal{F}_k^L$, where $k = 1, \cdots, N_c$. These ensembles are then aggregated to 
obtain the output uncertainty for the QoI. As mentioned briefly in the beginning of this section, the 
multi-model strategy for uncertainty quantification uses data from high-fidelity simulations to construct model discrepancy functions to correct low-fidelity model results. This procedure, which is not present in traditional MC methods with single-model approaches, can be considered implicit model calibration. This calibration differs from tradition calibration procedures in two major aspects: (1) it takes into account the errors of calibration data themselves (i.e., the high-fidelity simulation), and (2) it is probabilistic instead of deterministic, considering the interpolation uncertainties of the model discrepancy based on GP assumptions.

Three sampling procedures are required in the multi-model uncertainty propagation method outlined
above.  The first one, which is used to obtain design set $\mathcal{D}^L$, is the same as in traditional 
MC methods and does not need further discussions. In this work we use Latin hypercube sampling 
to perform this sampling. Another sampling is required to obtain design set $\mathcal{D}^H$. 
Since points in this ensemble are used to construct Gaussian processes, we use a deterministic, 
uniform sampling on the range of parameter domain to minimize the largest possible distance between 
an arbitrarily chosen point in the domain and the nearest sampled points. More sophisticated sampling 
schemes are possible~\cite{sacks1989design}, which will be pursued in future studies. Finally, 
correction step involves sampling a \emph{function} from the posterior GP as described in 
Eq.~(\ref{E:postGP}). In practical implementations, only the values of the function at the parameter 
locations corresponding to the samples in the ensemble $\mathcal{D}^L$ are needed. Therefore, we 
only sample $N^L$ random numbers with joint Gaussian distribution $\mathcal{N}(\mathbf{0}, K) $, 
where $K$ is the blocked kernel matrix in Eq.~(\ref{E:postGP}). This is performed with standard
procedures~\cite{williams2006gaussian}, i.e., a vector $\tilde{\mathbf{x}}$ consisting of $N^L$
independent and identically distributed (i.i.d) random numbers is generated first, and then a
transformation $\mathbf{x} = L \tilde{\mathbf{x}}$ is performed to obtain a random vector with
desired covariance $K$, where $LL^T = K$.  Eigenvalue decomposition is used to obtain $L$ from $K$,
whose computational cost scales as $O(N^3)$ with $N$ being the size of the random vector
$\mathbf{x}$. This can be expensive when $N$ is large. However, it is worth pointing out that the
sample does not involve numerical model evaluations. Various standard techniques are available to
reduce the computational expense of this operation, e.g., by partitioning the parameter space via
treed schemes~\cite{gramacy2008bayesian}.

\section{Numerical Simulations}
\label{S:Test} 
In this section we first evaluate the multi-model approach on synthetic 
cases that are of relevance to CFD applications. Using synthetic cases enables a comprehensive 
study on scenarios with a wide range of complexities. A realistic CFD case, the compressible flow 
over a NACA0012 airfoil, is then used to demonstrate merits of the multi-model approach for
uncertainty propagation.

\subsection{Synthetic Test Case}
\label{S:syn} 
\begin{figure}[htbp]
	\centering
	\includegraphics[width=0.5\textwidth]{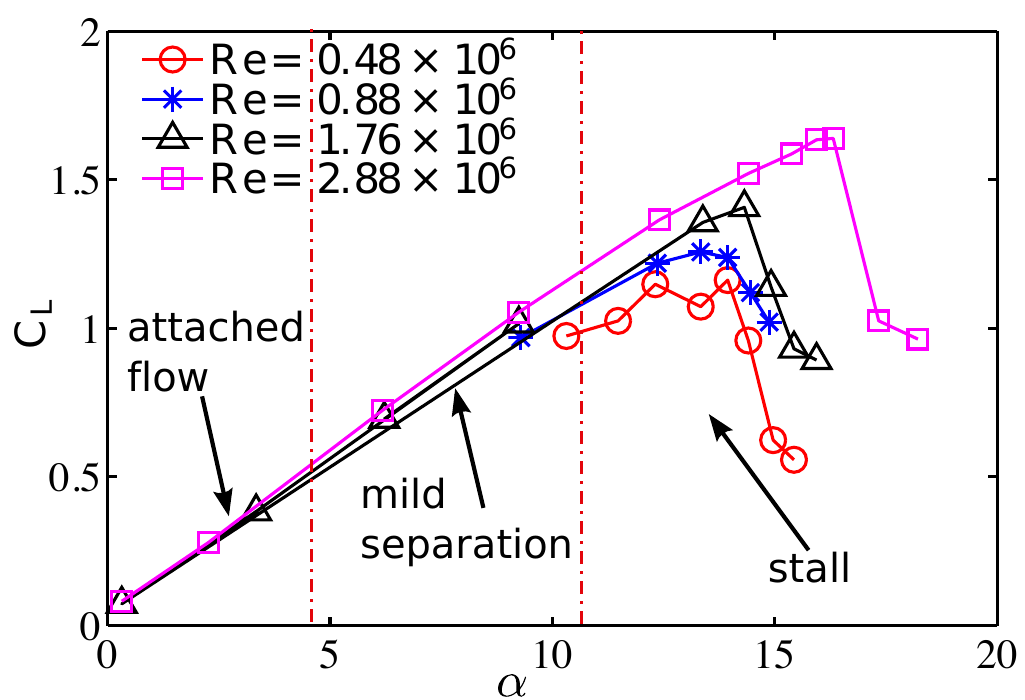}
	\caption{The dependence of the lift coefficient $C_L$ of a NACA-0012 airfoil section on AoA and Reynolds number $Re$. The three different flow regimes are demarcated with dash-dotted lines. }
	\label{fig:lift}
\end{figure}

When choosing test cases, we have the typical problem of airfoil lift coefficient prediction in
mind. In line with the example given above, one is interested in the uncertainty 
distribution of the lift coefficient $C_L$ when the operation conditions 
(i.e., input variables such as AoA $\alpha$, Reynolds number $Re$) 
conform to known distributions. The uncertainty propagation involves performing numerical 
simulations on a number of samples from the input (e.g., $x= [\alpha, Re]$) space to obtain 
samples in the output $C_L$ space. The simulations can be conducted by using a cheap, 
low-fidelity model such as panel method, or an expensive, high-fidelity model, or any models 
with computational costs and fidelities in between. The $C_L(\alpha, Re)$ response surface for 
a much studied classical airfoil, NACA-0012, is shown in Fig.~\ref{fig:lift}, which are fitted from
experimental data~\cite{gregory1973low}. A notable feature of this response surface is the smooth,
monotonic increase of $C_L$ with respect to AoA $\alpha$ in most regions and a sharp decrease at
certain AoA, corresponding to stall.  The flow physics and regime changes are dominated by the
AoA, but note that the Reynolds number provides modulations as it influences the angle at which the
stall occurs.

\begin{figure}[!h]
	\begin{center}
		\subfloat[ Monotonic]{\includegraphics[width=0.45\textwidth]{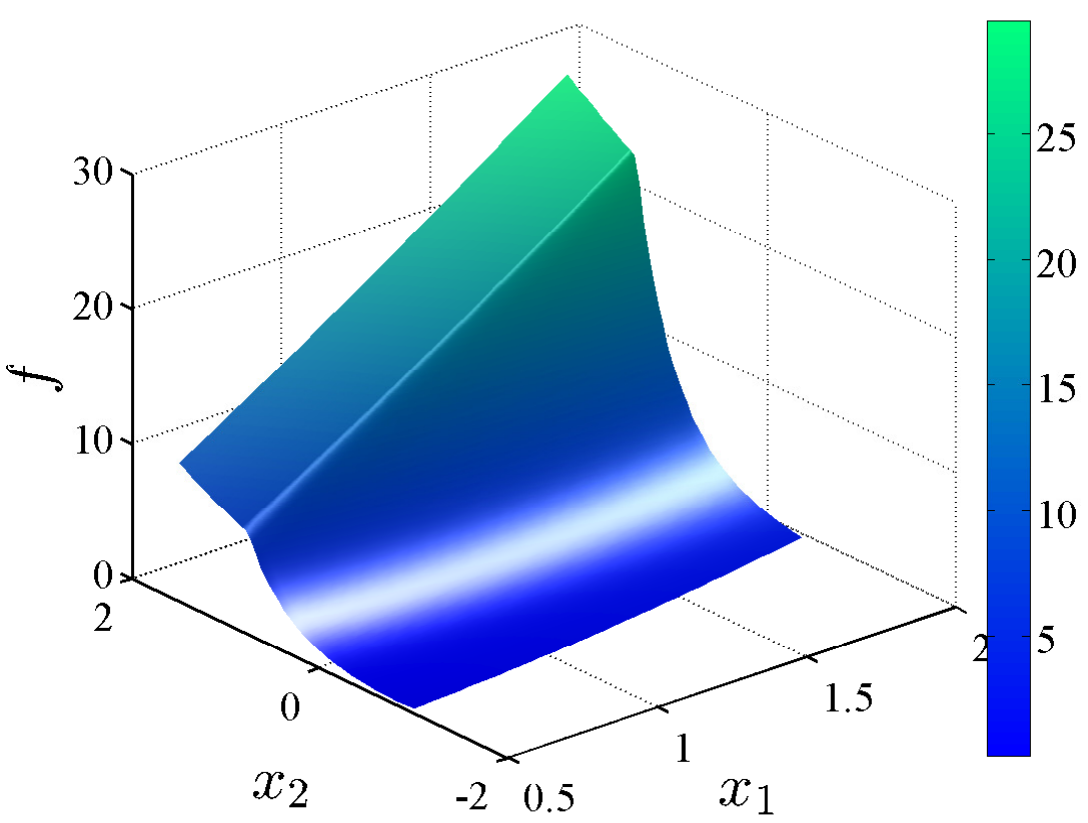}}
		\subfloat[Cross-sections Monotonic]{\includegraphics[width=0.45\textwidth]{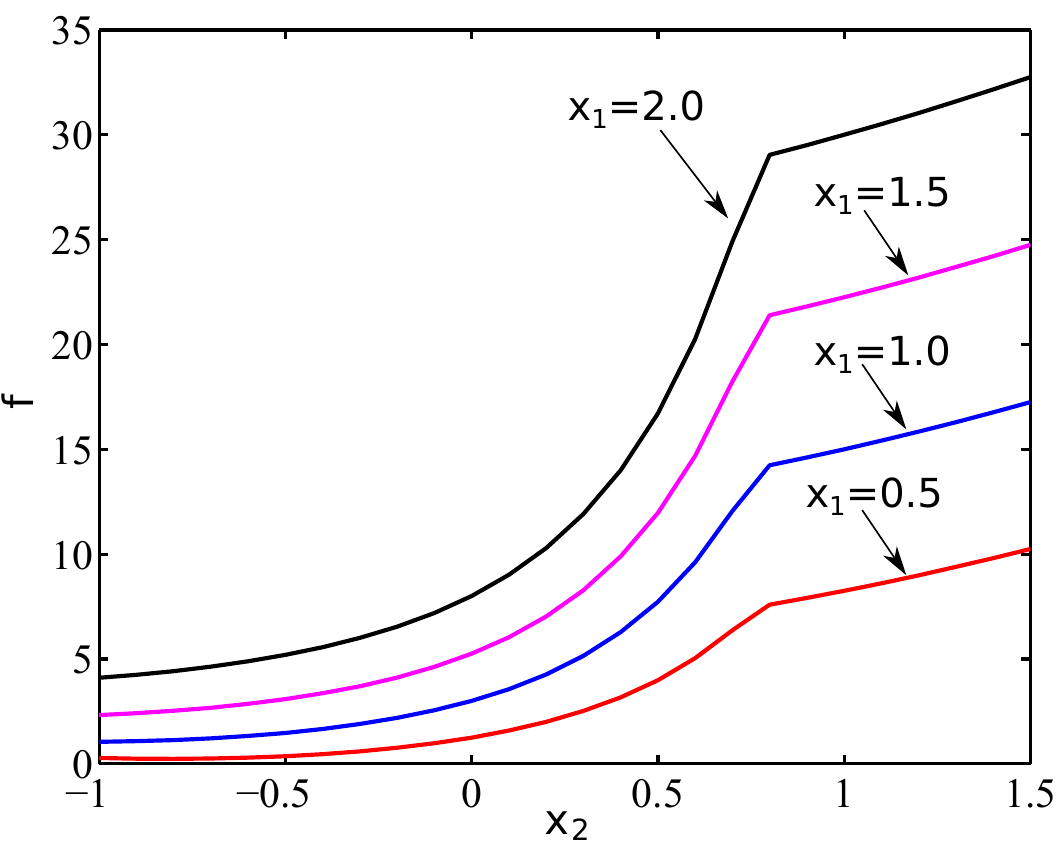}} 
	\end{center}
	\caption{
		\label{fig:MonSurf}
		Plots of the synthetic response surface  $f(x_1, x_2)$ in Eq.~(\ref{E:mon}). 
		The three-dimensional elevated surface with contour is shown in panel (a), and the cross-sections at several values $x_1$ are shown in panel (b).}
\end{figure}

For the purpose of evaluating the proposed method comprehensively, we first use a few 
synthetic response surfaces that closely mirror the features of the actual response surfaces 
in the airfoil example. Specifically, the response surfaces with different complexities 
from Helton and Davis~\citep{Helton:2003fc} are used. Due to the space limit, 
we only present the typical results of the first synthetic response surface in this work. 
For a comprehensive study on all synthetic cases, please refer to the electronic 
supplementary materials of this paper. The first response surface has the following expression:
\begin{equation}
\label{E:mon}
f_1(x_1, x_2) = x_1 + x_2 + x_1x_2 + x_1^2 + x_2^2  + x_1 \min(\exp(3x_2), 10). 
\end{equation}
which is a monotonically increasing function with a change of characteristics near plane 
$x_2 = 0.7$, as plotted in Figs.~\ref{fig:MonSurf}a and~\ref{fig:MonSurf}b. 
The specific locations of the characteristics change depend on the value $x_1$. In this mapping, 
$x_2$ dominates the feature of the response surface, and $x_1$ provides minor modulations, 
which is reminiscent of the roles of $\alpha$ and $Re$ in the airfoil example. One can compare Fig.~\ref{fig:MonSurf}b with Fig.~\ref{fig:lift} to appreciate this analogy. 

To mimic the behaviors of low-fidelity models (e.g., panel method or RANS solvers) and
high-fidelity models (e.g., LES) in CFD applications, the low-fidelity model is constructed 
by adding a discrepancy function $\delta(x)$ to the true response $f(x)$, with the discrepancy consisting of a constant $B$ and a part that is proportional to $f$, i.e.,
\begin{equation}
\label{E:lowfi}
z^L(x) = f(x) + \delta(x) \quad \textrm{with} \quad  \delta(x) = \nu  f(x) + B
\end{equation}
where $\nu$ is the proportionality constant, and in all cases, $\nu = 0.2$ and $B=2$ are used.
On the other hand, the high-fidelity model is constructed by adding an i.i.d. noise 
$\varepsilon(x') \sim \mathcal{N}(0, \sigma_n^2)$ to the true value. The variance 
$\sigma_n^2$ is chosen to be 0.01, which is much smaller than the low-fidelity 
model discrepancy $\delta$ in Eq.~(\ref{E:lowfi}). 

\begin{figure}[!h]
	\begin{center}
		\subfloat[Prior GP]{\includegraphics[width=0.45\textwidth]{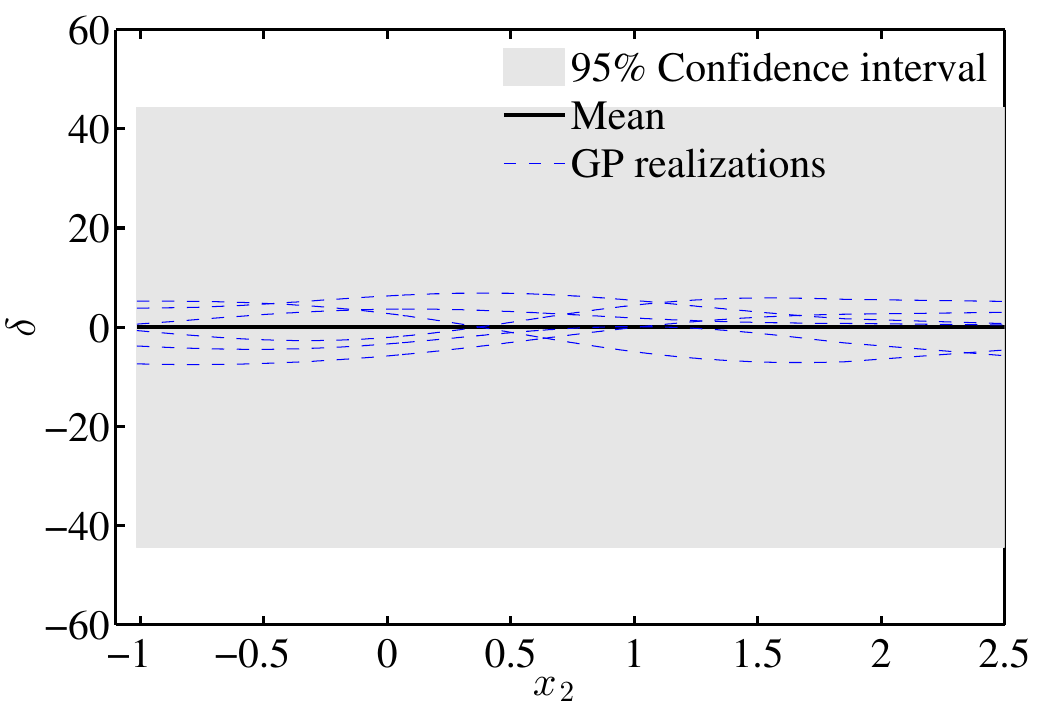}}
		\subfloat[Posterior GP]{\includegraphics[width=0.45\textwidth]{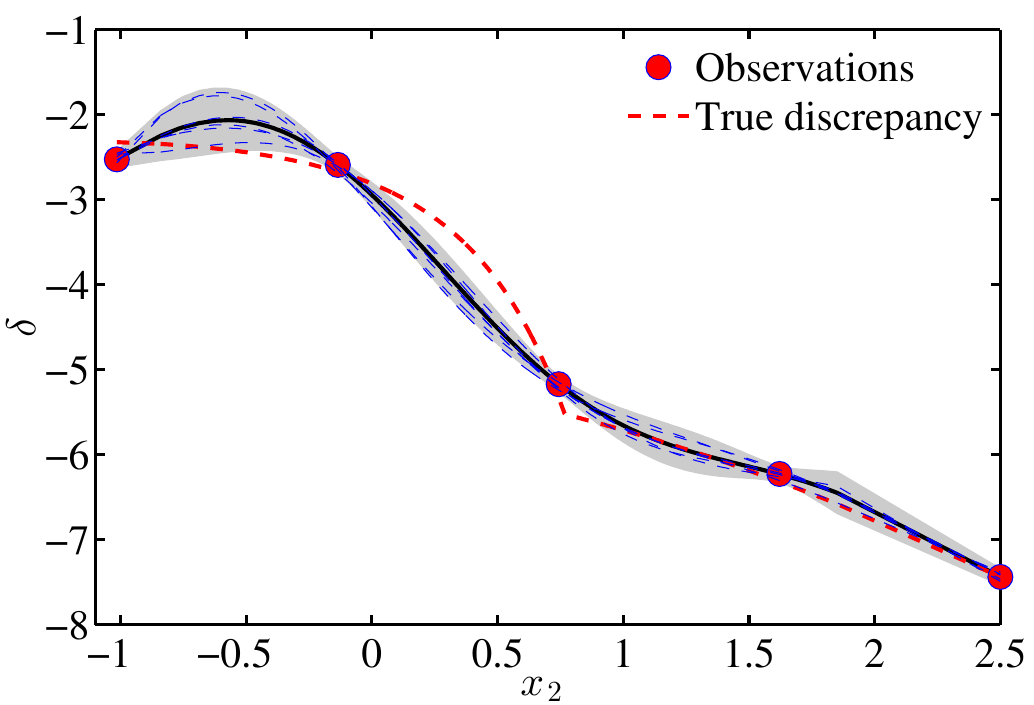}}\\
	\end{center}
	\caption{ (a) The prior distribution of the model discrepancy function $\delta$ is represented as a
		Gaussian process (GP) with a zero-mean function and a stationary kernel. (b) The posterior GP of
		$\delta$ given observation data at five input locations.  In both the prior and the posterior GPs,
		the mean functions, 95\% confidence intervals, and several random realizations from the GPs are
		indicated.}
	\label{fig:Mon1GP}
\end{figure}

Since the synthetic response has two input variables. To facilitate graphical presentations 
and to help understand the basic properties of the proposed method, we first test
the one-dimensional response surface, which is obtained by fixing the less important variable
as $x_1 = 1.25$. The response surface of model discrepancy is constructed with a GP model,
and the initial values for hyperparameters $\sigma$ and $l$ are 1.0 and 0.5, respectively. 
With five high-fidelity model evaluations, the assumed prior of the model discrepancy is updated 
by Bayesian approach to obtain the posterior GP. The prior and the posterior are presented in Fig.~\ref{fig:Mon1GP}a and~\ref{fig:Mon1GP}b, respectively.
Figure~\ref{fig:Mon1GP}a shows an uninformative prior with relatively large confidence intervals 
due to the lack of information. Given the ``observed'' model discrepancy data calculated from the
high-fidelity model results at the parameters in ensemble $\mathcal{D}^H$, the confidence interval
of the posterior GP is significantly reduced and the bias error has also been corrected (Fig.~\ref{fig:Mon1GP}b). It is noteworthy that the confidence interval is small (but not zero) close
to the observation points and increases further away from these points. In most part of the domain,
the confidence interval covers the truth, with the exception between $x_2 = -0.5$ and 0.8, where the
inferred distribution of $\delta$ becomes slightly over-confident. This is due to the fact the 
modular Bayesian approach, in which the hyperparameters are determined by the MLE method, 
only accounts for the most likely hyperparameter pair and ignores other less likely possibilities. 
This inevitably leads to overconfidence, particularly in the cases where the data are sparse, and 
thus many possibilities are equally likely. With this constructed model discrepancy GP, correction to 
the ensemble of 500 low-fidelity model results can be conducted by drawing $N_c$ realizations 
from the posterior GP.
 
\begin{figure}[htbp]
	\begin{center}
		\includegraphics[width=0.45\textwidth]{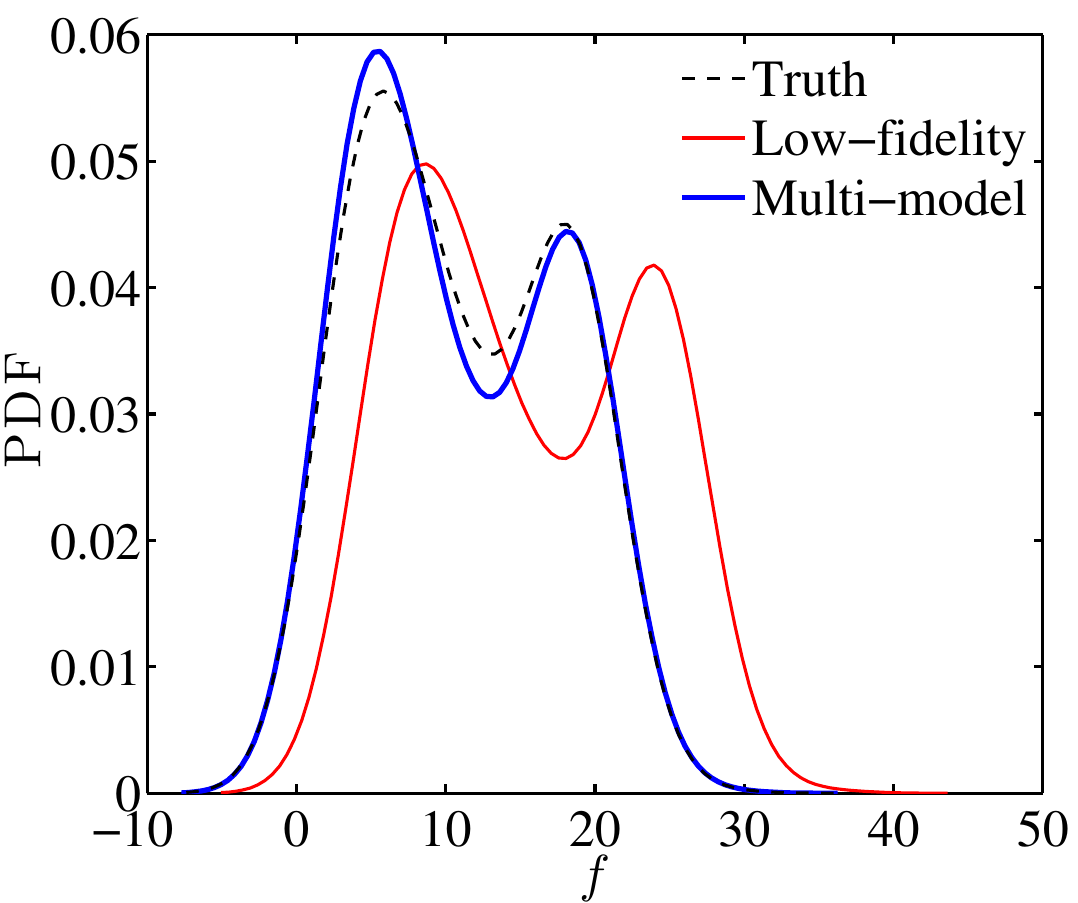}
		\caption{Corrected propagated uncertainty (blue bold solid line) based on the multi-model method
			compared with original results (red solid line) from the low-fidelity model. }
		\label{fig:Mon1DCDF}
	\end{center}
\end{figure}
The output ensembles after the correction are aggregated, and their Probability Density Functions
(PDF) are plotted in Fig.~\ref{fig:Mon1DCDF} with comparison to the truth and that obtained with the
low-fidelity model alone. Rigorous, quantitative assessment and comparison of predicted
distributions (in the form of PDFs or CDFs) are not straightforward and can be challenging. In
this proof-of-concept study we only conduct approximate comparison via visual observations, which
can be inadequate when the qualities of two predicted distributions are not obvious. In those cases,
more advanced metrics for comparing probability distributions are required, e.g., Area Validation
Metric~\cite{Oberkampf:2010uw} and its modified variant~\cite{voyles15evaluation:AIAA:15}. Figure~\ref{fig:Mon1DCDF} shows that the
normally distributed input uncertainty is mapped to a bi-modal distribution in the output
uncertainty distribution. Comparing results in this figure, it is clear that the PDF of the
corrected ensemble almost coincides with the corresponding truth, which indicates that the nonlinear
model discrepancy has been reconstructed and corrected reasonably well with only five data points
from high-fidelity simulations. A minor difference on the left peak and the valley of the PDF is due
to the slight overconfidence of the inferred posterior GP as discussed above. This simple case
demonstrates that the proposed multi-model scheme can effectively combine the results of models with multiple fidelities to improve uncertainty propagation process. We also evaluate the 
proposed uncertainty propagation scheme on the synthetic cases with other more complex responses surfaces used in~\cite{Helton:2003fc}, and qualitatively similar improvements are observed (see the electronic supplementary materials).

\subsection{Realistic Case in CFD Simulation of Flow Over Airfoil}
\label{sec:realCFDmapping}
\begin{figure}[!h]
	\begin{center}
	\subfloat[$C_L(\alpha, M_{\infty})$]{\includegraphics[width=0.47\textwidth]{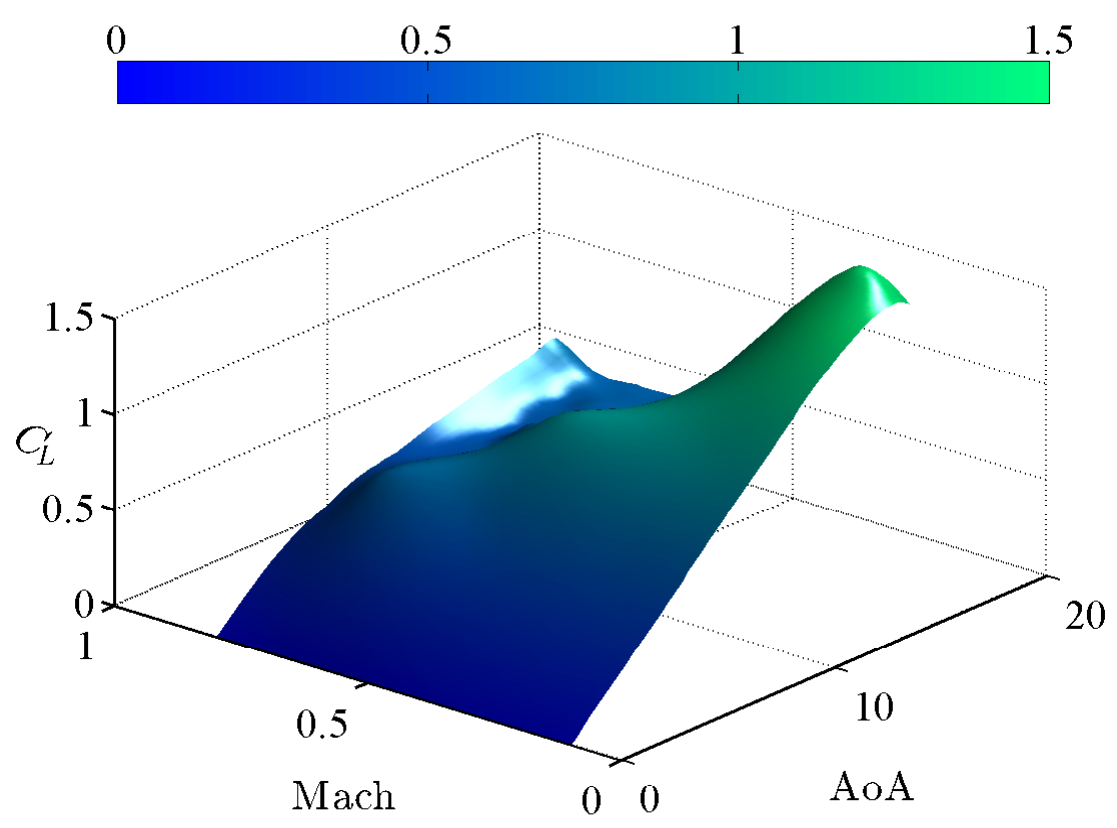}}
	\subfloat[$\delta$]{\includegraphics[width=0.45\textwidth]{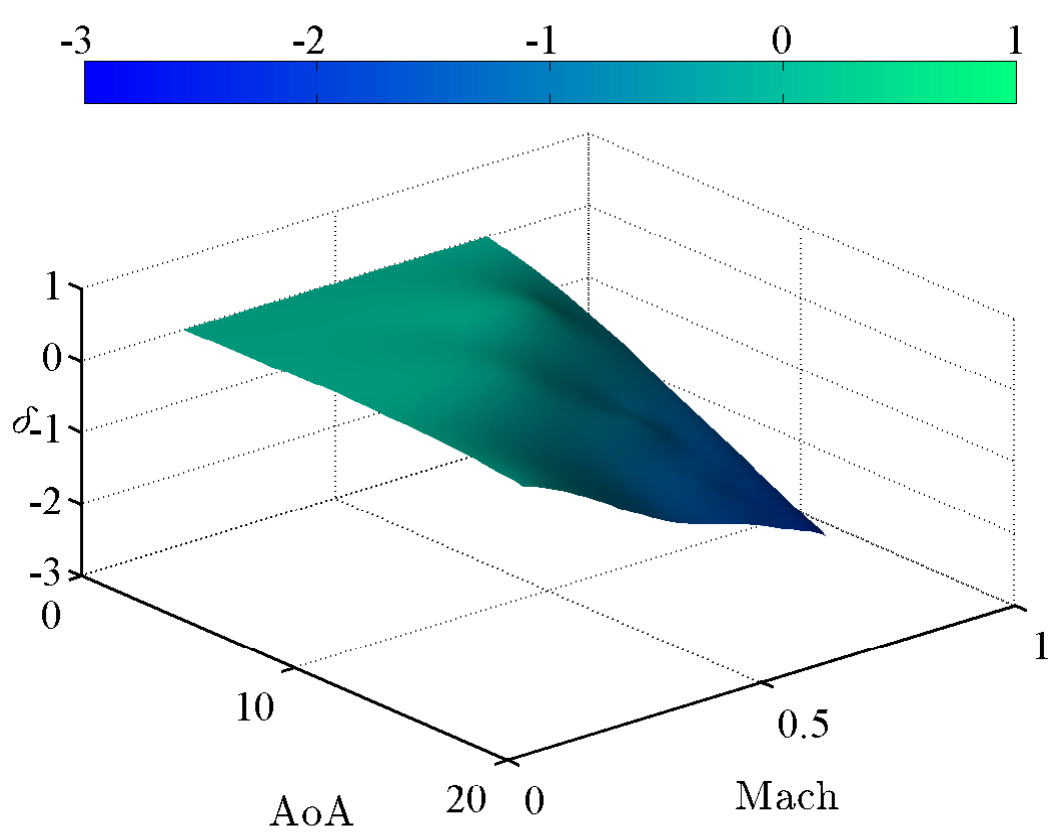}}
	\end{center}
	\caption{
		\label{fig:cfdResponse}
		Response surfaces of (a) the mapping $C_L(\alpha, M_{\infty})$ constructed by RANS solutions
		and (b) model discrepancy of low-fidelity model (i.e., thin airfoil theory with the 
		Prandtl--Glauert compressibility correciton). 
		}
\end{figure}

To illustrate the multi-fidelity uncertainty propagation approach on a realistic CFD case, 
a turbulent compressible flow over a NACA 0012 airfoil is used as a test example. 
The input parameters are angle of attack ($\alpha$) and Mach number ($M_{\infty}$), and 
the response is the lift coefficient $C_L$. In this example the response surface for 
$C_L$ as a function of $\alpha$ and $M_{\infty}$ is constructed from solutions of the 
steady RANS equations. Spalart-Allmaras one-equation turbulence model is employed to 
calculate the turbulent flow. The airfoil has a chord length of $c = 1$~m, and the simulations 
are conducted using a density-based solver with an ideal gas equation of state, a free-stream 
pressure of $101,325$~Pa, and a constant molecular viscosity of $\mu = 1.985 \times
10^{-5}$ kg/(ms). Adiabatic, no-slip walls boundary condition is imposed on the airfoil surface, and
free-stream far-field is applied to the outer boundary.  A matrix of RANS solutions with nine
different $\alpha$ and six different $M_{\infty}$ is used to build the ``truth'' response
surface. This two-dimensional response surface is constructed by the biharmonic spline fits, which
ensure the smoothness (see \cite{voyles2014evaluation} for details).  The true response surface for
$C_l$ with AoA and Mach number as inputs is shown in Fig.~\ref{fig:cfdResponse}a. It can be seen
that, similar to the experimental observations, the lift coefficient $C_l$ increases monotonically
with respect to AoA in most regions.  After certain threshold value of AoA, the lift coefficient
$C_l$ decreases markedly, which corresponds to the stall. When the Mach number is small, this
sharp decrease of $C_l$ at large AoA can not be observed. The chord-based Reynolds numbers 
for these Mach numbers are around $1 \times 10^7$.

In contrast to the synthetic case, we use a classic simplified aerodynamics model as the
low-fidelity model, where the lift coefficient is calculated based on thin airfoil theory along with
the Prandtl--Glauert compressibility correction~\cite{anderson1985fundamentals}:
\begin{eqnarray}
C_L^{(L)}(\alpha, M_{\infty}) = \frac{2 \pi \alpha}{\beta}\\
\beta = \sqrt{1 - M_{\infty}}
\end{eqnarray}
Similar to the synthetic case, the high-fidelity model of $C_L$ is constructed by 
adding an i.i.d. noise to the true value, i.e.,
\begin{equation}
\label{E:highfi}
C_L^{(L)}(\alpha, M_{\infty})  = C_L(\alpha, M_{\infty}) + \varepsilon,
\end{equation}
where $\varepsilon(x)$ is a white noise process with variance $\sigma_n^2 = 0.01$. The 
response surface of the model discrepancy of low-fidelity model $C_L^{(L)}(\alpha, M_{\infty})$ 
is shown in Fig.~\ref{fig:cfdResponse}b. As mentioned above, the true mapping $C_L(\alpha, M_{\infty})$ is obtained from the RANS simulations, while the low-fidelity model is based on the 
thin airfoil theory. It can be seen that the predictions from the low-fidelity model are relatively 
accurate when the AoA and the Mach number are small. This is because both thin airfoil theory 
and RANS solver can capture the flow feature and obtain accurate lift predictions in the region 
of attached flow. However, when the AoA or the Mach number increases, the flow separation 
occurs, which cannot be captured by the thin airfoil theory. Therefore, large discrepancy can 
be observed as the AoA or the Mach number is large. Similar to the synthetic case, the response
surface of discrepancy here also exhibits monotonic yet nonlinear dependences on the AoA 
and the Mach number.

 \begin{figure}[htbp]
 	\begin{center}
 		\includegraphics[width=0.45\textwidth]{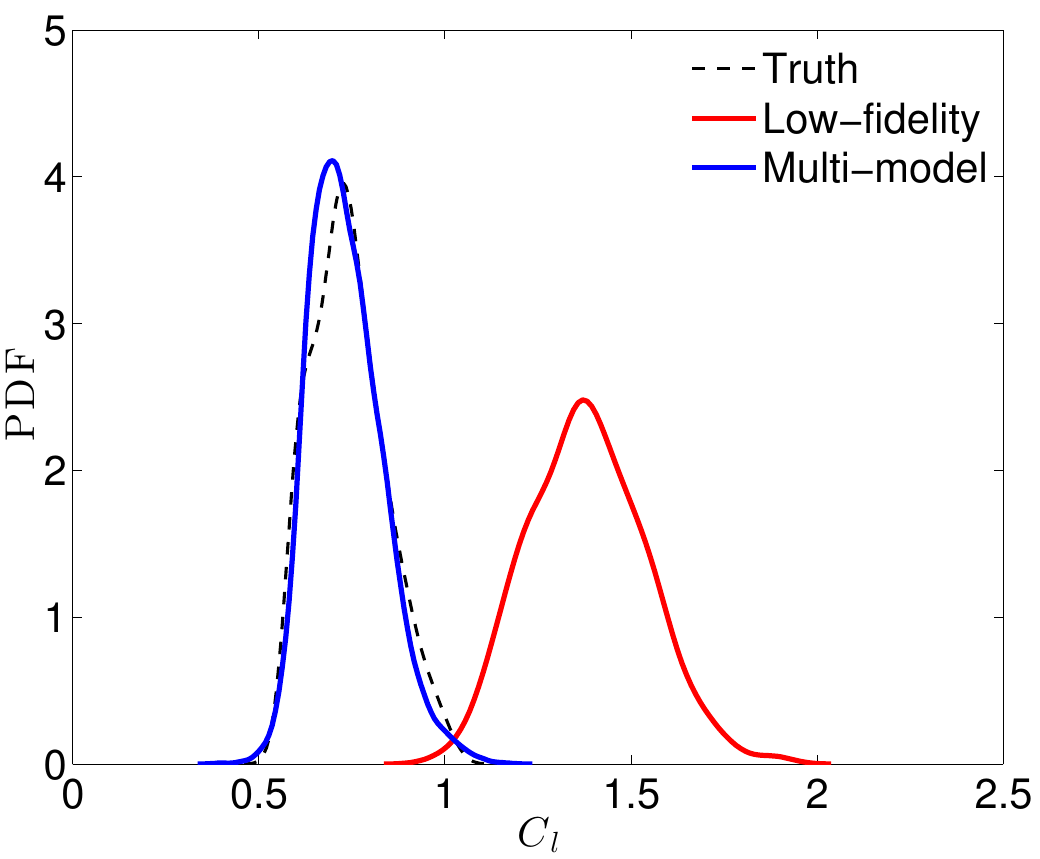}
 		\caption{Corrected propagated uncertainty (blue bold solid line) based on the multi-model method
 			compared with original results (red solid line) from the low-fidelity
                        model. Ten high fidelity simulations are used.}
 		\label{fig:CFD_PDF}
 	\end{center}
 \end{figure} 
In this case, two random input variables $\alpha$ and $M_{\infty}$ are assumed to be 
independent and normally distributed with $\alpha \sim \mathcal{N}(10, 1)$ and 
$M_{\infty} \sim \mathcal{N}(0.6, 0.06)$, respectively. Similarly, $N^L = 500$ samples
are propagated to the QoI, i.e., lift coefficient $C_L$. The PDF of propagated samples 
of $C_L$ is plotted in Fig.~\ref{fig:CFD_PDF} with comparisons to the truth and that 
obtained with the low-fidelity model alone. The normally distributed inputs ($\alpha$ 
and $M_{\infty}$) are propagated to $C_L$ (using the multi-fidelity model with 10 high
fidelity data), which is also normally distributed with a mean of approximately 0.7. 
However, the low-fidelity model over-predicts the lift coefficient and its distribution is 
distorted by the model errors. Similar to the synthetic case, PDF of the corrected ensemble 
nearly coincides with the corresponding truth, demonstrating the merits of the proposed 
uncertainty propagation approach.

 \begin{figure}[!h]
 	\begin{center}
 		\subfloat[10 high-fidelity simulations]{\includegraphics[width=0.45\textwidth]{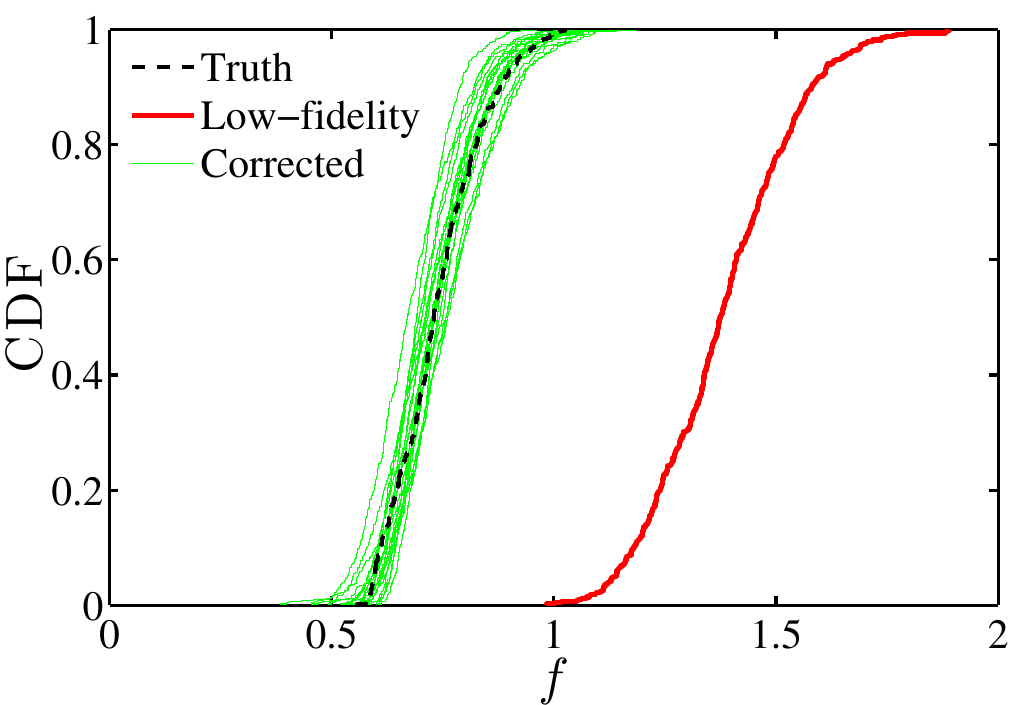}}
 		\subfloat[40 high-fidelity simulations]{\includegraphics[width=0.45\textwidth]{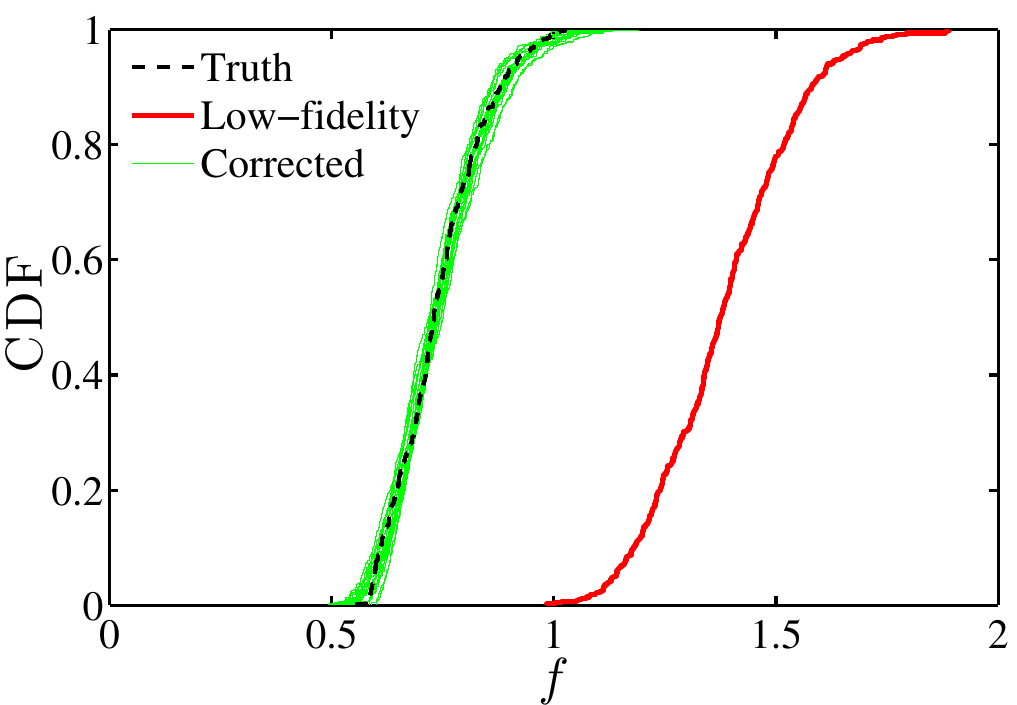}}
 	\end{center}
 	\caption{Individual CDFs of QoI corrected by using 30 realizations of the GP (before
 		aggregation) with (a) 10 high-fidelity simulations and (b) 40 high-fidelity simulations.}
 	\label{fig:cdfboxcfd}
 \end{figure}
To investigate the influence of the amount of available data on the propagated uncertainty, 
we compare the cumulative distribution functions (CDFs) of the $C_L$ ensembles 
corrected with 10 and 40 high-fidelity data points, where each CDF sample is corrected by an 
individual GP realization before the aggregation. The CDFs comparison is shown in 
Fig.~\ref{fig:cdfboxcfd}. For both cases, the probability regions covered by scattered CDF 
ensembles cover the truth and is significantly improved compared to the low-fidelity results. 
Note that we use a large number of individual GP realizations to conduct many corrections,
instead of using the predicted mean of GP to perform one deterministic correction. Hence,
an ensemble of corrected CDFs is obtained instead of a single CDF, which explicitly 
accounts for the uncertainties introduced by the lack of data. 
By comparing Fig.~\ref{fig:cdfboxcfd}a and.~\ref{fig:cdfboxcfd}b, we can see that the 
scattering of CDF samples with 10 high-fidelity data is larger than that with 40 
high-fidelity data points. This means the output uncertainty is enlarged, or stated
differently, the information entropy increases, when the data is reduced. 
This is consistent with what we expect that lack of data leads to increase of uncertainty 
in the propagated distribution, which is referred to as \emph{uncertainty inflation}.
In the proposed multi-fidelity strategy, the uncertainties introduced in the reconstruction 
of discrepancies are obtained based on GP assumptions, and they almost always inflate 
the propagated input uncertainty. This inflation can be reduced by using more information 
from high-fidelity simulations.   

\section{Discussion on Advantage Over Single-Model Approaches}
\label{S:diss}

When evaluating the merits of the proposed multi-model approach, there are two legitimate 
questions to ask before the proposed method can be justified. (1) What benefits does the 
high-fidelity model offer?  Or in other words, would it be better to simply allocate all 
computational resources to low-fidelity simulations?  (2) What benefits does the low-fidelity 
model offer? Or in other words, would it be better to allocate all computational sources to 
high-fidelity simulations?  Inevitably, answers to these questions are not straightforward, and 
ultimately they depend on the relative merits and costs of the respective models. Consequently, 
investigations of these issues can only be based on specific low-fidelity and high-fidelity models.  
In the following we will discuss the two issues above based on the models assumed in synthetic case. 

\begin{figure}[htbp]
	\begin{center}
		\includegraphics[width=0.45\textwidth]{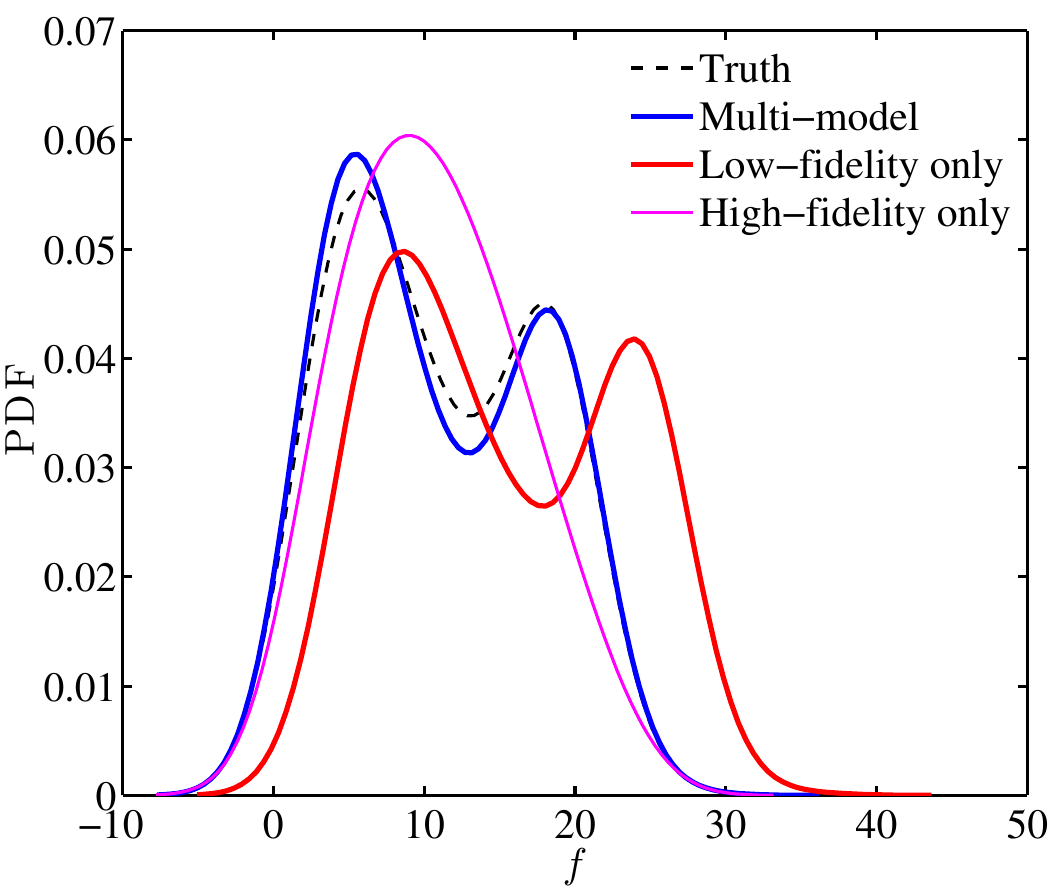}
		\caption{Propagated uncertainty distribution obtained by using the proposed multi-model method
			compared with those obtained by using the single-model approaches, i.e., with either high-fidelity
			model or the low-fidelity model alone.}
		\label{fig:pdfDis}
	\end{center}
\end{figure} 
The first question on the benefits of the high-fidelity model has been answered clearly in
Section~\ref{S:Test} .  When the low-fidelity model alone is used in the uncertainty propagation,
the obtained output uncertainties show significant biases, which are due to the prediction bias 
of the low-fidelity models. In contrast, when data from high-fidelity simulations are incorporated 
into the uncertainty propagation process, even though the amount of available data is very limited, 
the improvements are clearly visible. This observation effectively demonstrates the value of the 
high-fidelity model, without which the correct output uncertainty cannot be obtained, no matter 
how many samples are used in the low-fidelity model evaluations. This is illustrated in Fig.~\ref{fig:pdfDis}, where the PDF obtained by using the low-fidelity model alone shows a 
significant bias even with a very large ensemble with 500 samples. The second question is 
closely related to the investigations conducted by Higdon et al.~\cite{higdon2004co} regarding 
the usefulness of numerical simulations in a prediction method combining numerical simulations 
and experimental data. They concluded that the numerical simulations have significant contributions 
in reducing the prediction uncertainties, and thus the combined simulation--data based framework 
is superior to the pure experimental data-driven approach. Here we carry out a similar investigation 
in the context of uncertainty propagation by comparing the multi-model results with those obtained 
by utilizing high-fidelity simulations alone. That is, in this alternative approach, data obtained from 
high-fidelity simulations are used to fit a GP, from which samples are drawn to propagate the input
uncertainties. The low-fidelity-mode is not used in this alternative approach. Results from the 
simplest one-dimensional response surface from synthetic case discussed in
Sec.~\ref{S:syn}, are presented to illustrate the benefits of the low-fidelity model. For this 
comparison, five high-fidelity model results are adopted for both methods and the same
parameters as above are used. The PDF of the QoI uncertainty distribution obtained by using the
multi-model approach and that obtained by using the high-fidelity model alone are shown in
Fig.~\ref{fig:pdfDis}. From this figure, one can see that the multi-model results agree with the
truth extremely well. In contrast, the results from the high-fidelity model alone deviate
significantly from the truth. Although the high-fidelity model based approach indeed corrected 
the obvious bias that is present in the pure low-fidelity model results, it fails to capture the
bi-modal feature of output uncertainty distribution, which is an essential feature present in the
true QoI distribution. Overall, the multi-model method gives much better results for the QoI
uncertainty than the single-model approach based on the high-fidelity model alone.
In summary, the comparison above suggests that, by combining the low- and high-fidelity 
models, the proposed approach produces better results than either model can yield alone.

\section{Conclusions}
\label{S:Con}
Model uncertainties play an important role in computational fluid dynamics and particularly in
turbulent flow applications, where first-principle-based direct simulations are prohibitively
expensive and thus are not affordable in most practical applications. In CFD applications, when
propagating input uncertainties to QoI uncertainty distributions, uncertainties due to the model 
used in the uncertainty propagation can distort the obtained output uncertainty. The effects
of the model uncertainties on the results should be properly identified, while, at the same time,
sufficient samples should be used to avoid sampling errors. In this work we propose a multi-model
uncertainty propagation strategy, where model uncertainties are accounted for by using numerical
models of different fidelities. Compared to previous work, the proposed method is based on a 
smaller number of assumptions. In particular, Gaussian processes are not assumed for the 
low- and high-fidelity models themselves and only for the model discrepancy, which is more 
realistic for practical CFD applications. Synthetic cases related to CFD applications and a realistic
CFD problem are used to demonstrate the merits of the proposed method. Simulation results 
indicate that biases and distortions of the propagated output uncertainty can be significantly 
reduced by incorporating data from the high-fidelity simulations through the multi-model 
framework. Inflation in the propagated output uncertainty associated with the model uncertainty 
is quantified (see discussion in Sec.~\ref{sec:realCFDmapping}). By incorporating more data from the high-fidelity simulations, the uncertainty 
inflation can be reduced (see Fig.~\ref{fig:cdfboxcfd}).  Overall, the results clearly demonstrate that, by combining the low- 
and high-fidelity models, the multi-model uncertainty propagation strategy leads to significantly 
improved results compared with what either model can achieve individually.








\end{document}